# Voice Analysis for Stress Detection and Application in Virtual Reality to Improve Public Speaking in Real-time: A Review

Arushi

James Cook University, Singapore 387380, arushi@my.jcu.edu.au

Roberto Dillon

James Cook University, Singapore 387380, roberto.dillon@jcu.edu.au

Ai Ni Teoh

James Cook University, Singapore 387380, aini.teoh@jcu.edu.au

Denise Dillon

James Cook University, Singapore 387380, denise.dillon@jcu.edu.au

Stress during public speaking is common and adversely affects performance and self-confidence. Extensive research has been carried out to develop various models to recognize emotional states. However, minimal research has been conducted to detect stress during public speaking in real time using voice analysis. In this context, the current review showed that the application of algorithms was not properly explored and helped identify the main obstacles in creating a suitable testing environment while accounting for current complexities and limitations. In this paper, we present our main idea and propose a stress detection computational algorithmic model that could be integrated into a Virtual Reality (VR) application to create an intelligent virtual audience for improving public speaking skills. The developed model, when integrated with VR, will be able to detect excessive stress in real time by analysing voice features correlated to physiological parameters indicative of stress and help users gradually control excessive stress and improve public speaking performance.

CCS CONCEPTS • **Human-centred computing** → **Human computer interaction (HCI)** → **Virtual reality** • **Computing methodologies** → **Machine learning** → **Machine learning algorithms** → **Feature selection** → **Real-time simulation** → **Artificial intelligence** → **Intelligent agents**

**Additional Keywords and Phrases:** Virtual reality, Affect sensing and analysis, Nonverbal signals, Real-time feedback system, Voice analysis, Intelligent virtual agents, Signal processing

## 1 INTRODUCTION

Public speaking is considered a vital skill [Schreiber 2011]. It has several benefits to individuals at various levels, such as rapport building and networking in professional as well as social contexts. The predominant functional state of speakers during public speaking is the experience of stress [Koroleva et al. 2014]. Several studies found that public speaking causes and induces excessive stress in speakers and therefore compromises the quality of speech being delivered [Beatty and Behnke 1991; Dowd et al. 2010; Elfering and Grebner 2011; Elfering and Grebner 2012; Garcia-Palacios et al. 2002; Georgiades et al. 2000; Gramer and Saria 2007; Gramer and Sprintschnik 2008; Kotlyar et al. 2008; Pisanski et al. 2016; Westenberg et al. 2009; Wiemers et al. 2015; Wirtz et al. 2013]. According to the American Psychological Association (APA), stress is defined as a physiological or psychological response to any internal or external stressor [American Psychological Association 2022]. Stress manifests in every system of the body, for example, through heart palpitations, dry mouth, shortness of breath, accelerated speech, and changes in breathing patterns leading to changes in voice [American Psychological Association 2022]. Although a moderate level of stress is beneficial for optimal performance, high levels of stress lead to fatigue and reduced performance [Dongrui et al. 2010]. Stress during public speaking not only impacts speaking ability in large social gatherings, but it also affects daily interpersonal communication skills. Ultimately, it impacts opportunities for learning and growth.

Despite public speaking commonly triggering stress among individuals, it has received relatively little attention in the fields of affective computing and human-computer interaction (HCI). This is possibly due to the limitations and complexities of creating a suitable testing environment. Specifically, little attention has been given to the real-time identification of stress during public speaking. Automatic detection of stress from audio and voice is one of the most challenging tasks because emotions (such as anxiety) and stress are basically indistinguishable in terms of reaction patterns and response to an event [American Psychological Association 2022].

In the past, researchers have tested various classification algorithms to analyse human voice for the identification of six basic, distinct categorical emotions, such as happiness, anger, sadness, surprise, disgust, and fear, and other dimensional emotions, such as anxiety, boredom, calmness, and sarcasm [Bitouk et al. 2010; Deng et al. 2013; Kwon et al. 2003; Lin and Wei 2005; Poorna et al. 2018; Soltani and Ainon 2007; Zhou et al. 2001]. Emotions are fundamental in human nature and impact both verbal and non-verbal vocalisations. Therefore, emotion classification algorithms can be developed by creating emotionally significant events which induce physiological, psychological, and complex behavioural and reactive patterns [American Psychological Association 2022]. The algorithms discern emotions using voice analysis and extracting specific features that are indicative of affective states. For instance, some studies [Ang et al. 2002; Liscombe et al. 2005; Lugger and Yang 2007] utilised prosodic features of voice which are essentially non-verbal patterns of tune in speech, such as speaking rate, pitch, energy and pause duration to help detect the speaker's state beyond the literal spoken words. Other studies [Rong et al. 2009; Sanchez et al. 2010; Schuller et al. 2007; Wang et al. 2015] used spectral features which are obtained from the time domain signals (temporal features) through the application of Fourier transforms to describe the signal in the frequency domain.

The process of feature extraction (e.g., prosodic or spectral) for detection/identification of specific emotional/affective state by voice analysis can be a difficult task considering both the event and the environment. Specifically, identifying features to detect stress in real-time during the event of public speaking is a complex and challenging problem to be solved, as the associated physiological changes, such as increased heart rate and blood pressure, are not exclusively indicative of stress experience [Clinton et al. 2017; Kappas et al. 1991; Westenberg, Bokhorst, Miers, Sumter, Kallen, van Pelt and Blöte 2009] Therefore, stress detection should include changes in voice and breathing patterns [Droppleman and McNair 1971]. Stress during public speaking can be reduced by participating in training programs as well as receiving social support [Fredrick et al. 2018; Teoh and Hilmert 2018; Thorsteinsson and James 1999; Uchino and Garvey 1997]. Furthermore, stress can be detected by organising speaking tasks in laboratory environments, for example, by using recent exponential technologies like Virtual Reality (VR). Application of the latest technologies like VR may help in overcoming limitations and complexities involved in creating an environment to simulate public speaking. Being a human-computer interface designed to immerse users in a simulated environment [Azwar et al. 2016], VR provides a safe space for HCI for various purposes, such as training [Fruchter et al. 2007] and applications in different industries. Since VR has the capability to engineer stress responses to create a more confident public speaker in real life situations, it could therefore be utilised as a training tool to help reduce excessive stress of public speaking in real-time.

This literature review aims to give readers a general overview of the various algorithms used to identify different affective states by analysis of various voice features. In this paper, our key contribution is as follows:
- We emphasise and present how a computational algorithmic model that specifically detects stress in real-time from voice analysis could be developed for improving public speaking. Its innovative features will include analysis of prosodic features such as intensity, frequency, the pitch of voice, and spectral features which are identified based on correlation established with physiological symptoms of stress such as heart rate, blood pressure and psychological responses obtained from questionnaires (Section 3).
- Next, we discuss how to implement (theoretically) a real-time stress detection model in VR for additional user immersion and engagement. Thereby, overcoming the limitation of testing environment for improving public speaking skills by effectively and efficiently reducing the gap between affective computing and human-computer interaction fields (Section 4).
- Additionally, virtual audience in VR can be trained using reinforcement learning techniques or other ways to detect stress and provide feedback in real time. The final developed VR application will include a virtual audience enabled to provide virtual social support in two different modes: supportive feedback mode where only positive feedback will be provided regardless of performance (measured in terms of detected stress levels), or realistic positive or negative feedback based on actual performance (Section 5). This will allow the speaker to practice techniques designed to reduce stress during public speaking first in a VR environment and then progressing on to refining public-speaking in real-life.

## 2  ARTICLE SEARCH AND SCREENING PROCESS

Numerous studies have examined the relationship between the voice and emotions. This review primarily focused on emotion detection from voice and VR for public speaking. Therefore, peer-reviewed articles and book chapters that refer to emotion/affective state detection, public speaking and voice analysis were explored by using ScienceDirect, Web of Science and IEEE Xplore databases. Only English-language publications were selected. Several different searches were conducted for the different topics using various keywords and combinations of keywords.

We conducted the first round of literature search (Search 1) to understand and identify the relationship between emotional/affective state and its effect on voice. The search included keywords *voice and emotion*, *public speaking*, and *virtual reality*. The setting of the timeframe for 50 years yielded important earlier research papers such as by Scherer [Scherer 1995], Wallbott & Scherer [Wallbott and Scherer 1986], and Kappas [Kappas, Hess and Scherer 1991], which were helpful in the basic understanding of the relationship between voice and emotional/affective state in humans. Similarly, some of the research conducted by Slater and his colleagues [Slater et al. 2006; Slater et al. 1999] helped elaborate the earlier research methods and experiments related to public speaking in virtual reality. Search 1 resulted in a total of 686 articles. These articles provided us with the necessary introduction to public speaking, voice, emotions, virtual social support, and virtual reality.

Before we made our Search 2, we were fairly confirmed in finding that little research has been conducted to detect stress in real-time by voice analysis. Therefore, our Search 2 focused on speech emotional/affective state recognition models and included keywords *audio and voice analysis*, *speech emotion detection* and *voice stress detection*. A timeframe of 20 years (i.e., 2000 – 2020) was set for Search 2 to understand the recent research related to affective state detection from voice. This was done mainly to get an overview of application and improvements in emotional state recognition models in the past 20 years. This provided more insights into changes that came in by using machine learning technology for emotion detection. Search 2 resulted in 438 articles.

Items from both searches were screened by using EndNote software [The EndNote Team 2013] was used to manage the references. After a screening of the total 1,124 abstracts and conclusions, 241 articles were selected. Articles which were less than a single page were eliminated, as were commentaries and reflective viewpoints. Furthermore, short papers with no data were excluded. Numerous articles which were not relevant and coherent with our research area were also removed. Research articles without a focus on emotional/affective state detection models, voice analysis features and VR for public speaking were excluded. Taken together, all search and screening methods resulted in a total of 169 articles.

## 3 LITERATURE SURVEY FINDINGS

### 3.1 Voice Analysis

Human voice is capable of conveying emotions through speech [Banse and Scherer 1996; Darwin and Prodger 1998; Scherer et al. 1991; Wallbott and Scherer 1986] and non-speech vocalisations [Cowen et al. 2019]. Furthermore, analysis of non-speech vocalisations (voice) and speech has received much recognition during the past decade to detect various emotional/affective states [El Ayadi et al. 2011; Schuller et al. 2011]. The state detection process from voice involves various algorithms which require training to work correctly.

An algorithm is a well-defined set of instructions that maps an input to an output using a function. For instance, if a given input is X (e.g., voice features) and we want to know if the output is Y (e.g., specific emotion/affective state), then the function involves mapping the given input X to Y as f(X)=Y. Detecting emotional/affective state via voice analysis also involves some preliminary steps such as pre-processing, signal framing, windowing, detection of voice activity, normalisation, and noise reduction of the data [Hamid 2018]. These steps play an essential role in extracting the specific features of voice capable of identifying different affective states and training the algorithmic models for classifying and recognising emotions.

**3.2 Voice Features and Emotion Detection**

The communication of emotions in humans takes place beyond the literal spoken words, such as in non-verbal patterns of rhythm, timbre, and tone in speech. Overall, many non-lexical patterns can be identified by features such as intensity, fundamental frequency ($F_0$) and duration of the sounds. The prosodic features, which are a combination of spectral, also identified as frequency-based features and including time-domain features, also known as temporal features, are of paramount importance for detecting affective states have also been discussed in the literature [Mitchell and Ross 2013].

Temporal features include intensity, which signifies the energy of the sound waves (signal) travelling through the air. It is the representation of a change in the amplitude of speech signals over time. Amplitude here in this process is the displacement of the sound waves produced, which determines the loudness of the voice [Kappas, Hess and Scherer 1991]. These features are relatively simple to extract and are easier to interpret than spectral features.

Spectral features specifically include features associated with the fundamental frequency ($F_0$) such as frequency components, formants, and coefficients. Vibrations of the vocal cords cause $F_0$, which denotes the lowest oscillation of sound waves to be produced, and it is related to the pitch of voice. $F_0$ one of the most-used parameters for detecting emotions [Scherer 1995]. The pitch comprises characteristics related to rhythm and tone of the speech. The variation of the $F_0$ over time yields its other statistical properties and $F_0$ contour, which can also be used as features [Kamiloğlu et al. 2020]. More specifically, formants indicate the location of resonance occurring in any vocal tract. They are particularly described with their specific frequencies, their amplitude at the peak, and their bandwidth. The localisation of these formants' resonance on frequencies causes amplification or attenuation. The resonance of the formants regulates articulated sound quality and are assessed as the peak in the frequency spectrum of the sound [Koolagudi and Rao 2012]. Spectral features are obtained by converting the time-based signals into the frequency domain signals using a Fourier Transform [Kappas, Hess and Scherer 1991; Scherer 1995].

One of the essential spectral features is the Mel Frequency Cepstral Coefficients (MFCC), which represents the speech signal's short term power spectrum. It is obtained by division of the utterances into segments. Then a short-time discrete Fourier Transform is applied on each segment to convert it into the frequency domain. Next, the Mel filter bank is utilised for the calculation of the number of sub-band energies, which is then followed by taking logarithms and finally the application of inverse Fourier Transform obtains MFCC [Kuchibhotla et al. 2014]. Other spectral features include Linear Prediction Cepstral Coefficients (LPCC), Log-Frequency Power Coefficients (LFPC) and Gammatone Frequency Cepstral Coefficients (GFCC) that can be acquired by a Linear

Prediction Coefficient (LPC), by measurement of spectral band energies using Fast Fourier Transform and Gammatone filter-bank respectively [Nwe et al. 2003].

Furthermore, an algorithm documented by Kaiser [Kaiser 1993] as

$$\Psi[X(n)] = x^2(n) - x(n+1)x(n-1)$$

was also developed to measure the energy from speech by a non-linear process, where $\Psi[]$ is the Teager Energy Operator (TEO), and $x(n)$ represents the sample of a speech signal. TEO has various features, such as TEO decomposed FM (frequency modulation), TEO decomposed FM variation (TEO-FM-Var), and TEO auto-correlation envelope area (TEO-Auto-Env), as well as a critical band based TEO auto-correlation envelope area (TEO-CB-Auto-Env). These features explore the change in the characteristics of energy's airflow of speech under stress [Sun and Moore 2011]. These features can be particularly utilised for detection of stress.

Another set of features called prosodic features are a combination of both temporal and spectral features. For instance, intensity, pitch, $F_0$, formants and duration are also categorised as prosodic features. Duration is defined as the time to construct vowels and other factors present in speech, such as words. Speech pace, the time duration of voiced, unvoiced, and silenced region and longest voiced speech are some of the commonly used duration related features. Furthermore, some other properties of a voice include shimmer, jitter and harmonics to noise ratio (HNR). These features might also help differentiate the emotions produced in a speech signal. While jitter represents the unevenness of $F_0$ between successive oscillations, the shimmer is the unevenness of the amplitude. Furthermore, the HNR is the representation of the ratio between periodic to aperiodic components in a voiced speech signal [Li et al. 2007]. In Table 1, we summarise common voice features used in emotion/affective state detection.

TABLE 1. Different types of voice features

| | | |
|---|---|---|
| **Temporal features (Time based features)** | Features pertaining to intensity | Minimum energy, zero-crossing rate, maximum amplitude |
| **Spectral features (Frequency based features)** | Represents the signal in terms of frequency | Fundamental frequency ($F_0$), formants, pitch, fundamental frequency contour |
| **Mel Frequency Cepstral Coefficients (MFCCs)** | Represents short term power spectrum | Linear Prediction Cepstral Coefficients (LPCC), Log-Frequency Power Coefficients (LFPC) and Gammatone Frequency Cepstral Coefficients (GFCC), Linear prediction Coefficient (LPC) |
| **Teager Energy Based features** | Represents energy in signal by following a non-linear process | TEO decomposed FM (frequency modulation), TEO decomposed FM variation (TEO-FM-Var), and TEO auto-correlation envelope area (TEO-Auto-Env), as well as a critical band based TEO auto- |
| | | correlation envelope area (TEO-CB-Auto-Env) |
| **Prosodic Features** | Combination of Temporal and Spectral features | Intensity, pitch, $F_0$, formants, duration |
| | | Unvoiced and silenced region, |
| **Some other voice features** | Represents unevenness of $F_0$ between successive oscillations | Shimmer |
| | Represents unevenness in amplitude | Jitter |
| | Represents representation of the ratio between periodic to aperiodic components in a voiced speech signal | Harmonics to noise ratio (HNR) |

It has been found that listeners can judge five types of emotions – anger, fear, happiness, sadness and sensitivity – with up to 70% accuracy [Juslin and Laukka 2003]. Researchers [Kappas, Hess and Scherer 1991] suggested that emotions such as anger, fear, anxiety, happiness or surprise, which are considered to be high arousal emotions, result in increased intensity as well as a high mean $F_0$. In comparison, disgust, and sadness, which are considered to be low arousal emotions, have the opposite effect. Kamiloğlu et al. [Kamiloğlu, Fischer and Sauter 2020] suggest that pitch is high for emotions such as amusement, interest, and relief, but moderate for savouring emotions like contentment and pleasure, and low for prosocial emotions such as admiration. Low et al. [Low et al. 2010] used prosodic, spectral voice quality as well as TEO-based features for detecting clinical depression in adolescents. The results showed that TEO-based features, specifically TEO CB AutoEnv, outperformed all other features and all other combinations of features as well.

Despite a range of voice features, most of the early studies in the last decade were focused on analysing the prosodic features including pitch, formant frequencies and intensity [Busso et al. 2009; Nogueiras et al. 2001; Nwe, Foo and De Silva 2003; Schuller et al. 2003]. Recent literature surveys showed extensive use of spectral features and cepstral measurements (e.g., MFCCs), voice quality features including HNR, jitter, or shimmer, and Teager energy operator such as TEO-FM-Var and TEO-CB-Auto-Env [Bhavan et al. 2019; Hao et al. 2020; Semwal et al. 2017]. This might be since various features perform differently in the detection of emotional/affective state based on the newly developed algorithms.

There exist various algorithms for analysing the voice and audio for emotion detection. The most significant ones are the Gaussian Mixture Model (GMM), the Hidden Markov Model (HMM), Support Vector Machines (SVM), Artificial Neural Networks (ANN), the Deep Neural Network (DNN) and the Convolutional Neural Network (CNN) [Hansen and Liu 2016; Julião et al. 2015; Přibil and Přibilová 2013; Schuller, Batliner, Steidl and Seppi 2011; Zeng et al. 2019].

### 3.3 Models for detecting Affective States from Audio and Voice Features

In recent years, the Gaussian Mixture Model (GMM) is one of the well-known models for emotion detection [Anagnostopoulos et al. 2015; Busso, Lee and Narayanan 2009; Tashev et al. 2017; Truong and van Leeuwen 2007]. GMMs are efficient in modelling multi-modal distribution that best model any input data, by finding probabilistic cluster assignments [Douglas-Cowie et al. 2007]. A GMM uses expectation – maximization [E-M] approach, where it chooses starting guesses for setting location and shape. It repeats the E-M steps until each cluster is connected with a smooth Gaussian model which can be in any form from stretched circular (Figure 1) to even oblong stretched out clusters (Figure 2) [VanderPlas 2016].

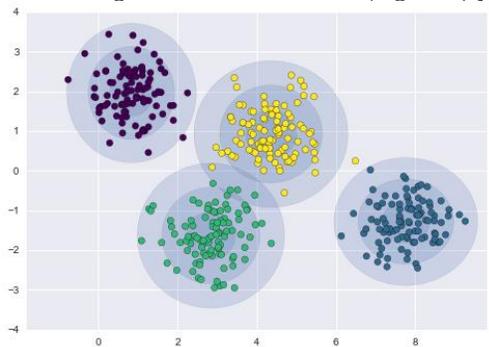
Figure 1: Circular stretched out clusters using GMM

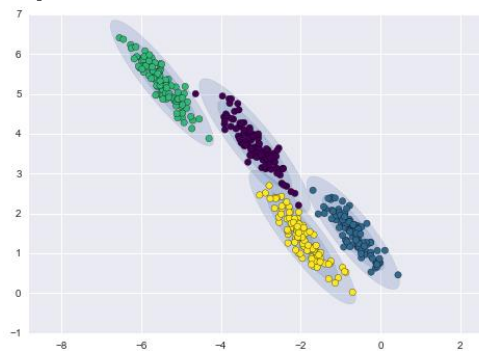
Figure 2: Oblong stretched out clusters using GMM

A group of researchers developed and tested the various GMMs with spectral features such as MFCC, MFCC-low, and pitch. Neiberg et al. combined three different models for recognising emotions. The developed models were tested on voice controlled telephonic service recordings recorded by Swedish company Voice Provider (VP) and on the meeting recordings from the ISL Meeting corpus. The results showed that the combination of developed GMMs based on the spectral features obtained the best results [Neiberg et al. 2006].

In another study, Schuller et al.[Schuller, Rigoll and Lang 2003] compared two models: GMM and HMM. GMM was used to classify the utterances using global statistics of raw pitch and energy contour. On the other hand, HMM classified the utterances using low-level statistical features (which represent a stationary state of emotions in voice). Each emotion in HMM is modelled by a single state HMM [Casale et al. 2007]. The single state HMM is trained by increasing the minimizing distance margin between emotions, and a loss function is used to scale the margin. In comparison to GMMs, the HMM is based on the Markov chain model and in its basic formulation, the Markov chain model states that the future state of any pattern is decided by the current state and not by the past state.

Markov Assumption:
$$P(q_i = a | q_1 ... q_{i-1}) = P(q_i = a | q_{i-1})$$
Where $q_1 ... q_i$ are the random i states [Martin. 2021].

It is similar to predicting tomorrow's weather based on today's weather only, without looking at the past weather. Furthermore, in HMM, the various states are kept hidden from the observer [Li, Tao, Johnson, Soltis, Savage, Leong and Newman 2007]. The hidden states of the HMM model indicate the temporal state of the voice data, where the temporal state is created because the observation sequence is formed due to its association with each state randomly. However, the calculation of the optimal number of states is an issue of HMM classifier design. The results indicated that the average accuracy in identifying discrete emotions was more than 86% by using global statistics on the features. Furthermore, GMMs are considered to be more appropriate for emotion detection as GMMs training and testing requirements are less than the continuous Hidden Markov Model (HMM).

Another HMM built model showed 70% recognition rate for identifying six emotions; happiness, anger, joy, fear, disgust, sadness; using low-level pitch, energy features of voice, and their contours of voice [Nogueiras, Moreno, Bonafonte and Mariño 2001]. Nwe et al. [Nwe, Foo and De Silva 2003] showed that the HMM yields better performance by achieving the best recognition rate of 89% by using spectral features MFCC and LPCC.

Similarly, HMM and Support Vector Machine (SVM) were used to classify five emotions: anger, happiness, sadness, surprise, and neutral emotion. The results from the study showed a 99.5% accuracy rate for HMM whereas SVM obtained an accuracy rate of only 88.9% in recognising the emotions [Lin and Wei 2005].

However, generally, in contrast to GMM and HMM, the SVM models are more recently used [Bhavan, Chauhan, Hitkul and Shah 2019; Bitouk, Verma and Nenkova 2010; Hao, Cao, Liu, Wu and Xiao 2020; Jain et al. 2020; Semwal, Kumar and Narayanan 2017; Truong and van Leeuwen 2007; Wang, An, Li, Zhang and Li 2015]. SVM seems to be promising in many research studies for the classification of stress or not stressed [Barreto et al. 2007]. It is used to find the best hyperplane for linearly separable patterns of any given data in an N-dimensional space (where N is number of features) which distinctly classify the data. Hyperplane here refers to the boundaries upon which a decision is made, whereby data falling on either side of the hyperplane can be considered with differed classes. Some of the possible hyperplanes for given data points are shown in Figure 3 [Gandhi 2018].

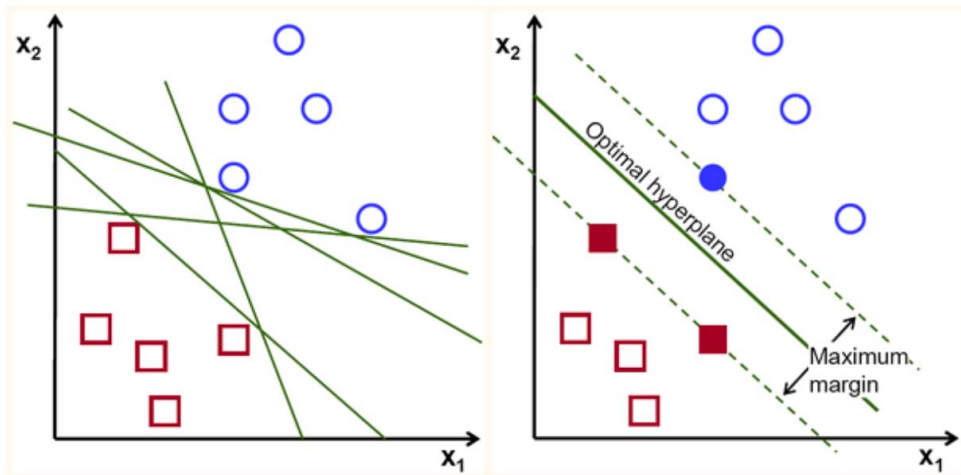

Figure 3: Possible hyperplanes

In case it is not possible to separate the patterns linearly, a kernel function is utilised to map the original data to a new space. However, since there is no particular defined way of choosing the kernel, the separation of transferred features is not guaranteed. Despite this, SVM also offers some specific advantages over the HMM and GMM such as global optimality of training algorithm and memory efficiency, since it uses a subset of training points in the decision-making function [Scikit Learn 2022].

Similarly, in the study conducted by Hu et al. [Hu et al. 2007] GMM supervector based on SVM for classification of emotions using spectral features reported an accuracy rate of 82.5%, while an accuracy rate of 77.9% was obtained by only using GMM.

Bhavan and colleagues [Bhavan, Chauhan, Hitkul and Shah 2019] recognised emotions such as happiness, fear, sadness, anger, disgust, boredom, calm, surprise, and sarcasm on three databases: the Berlin EmoDB [Burkhardt et al. 2005], the Indian Institute of Technology Kharagpur Simulated Emotion Hindi Speech Corpus (IITKGP-SEHSC) [Koolagudi et al. 2009], and the Ryerson Audio-Visual Database of Emotional Speech and Song (RAVDESS) [Livingstone and Russo 2018]. A combination of spectral features was extracted, further processed, and reduced to have the required features set including MFCC, spectral centroids, and MFCC derivatives. The group of researchers then further utilised a bagged ensemble model consisting of SVM with a Gaussian kernel. The results showed an improved accuracy rate of 92.45% (EmoDB), 75.69% (RAVDESS), and 84.11% (IITKGP-SHSC) respectively. Similarly, other researchers [Truong and van Leeuwen 2007] also used SVM for emotion recognition and found that SVM performs better with acoustic and linguistic features ultimately increasing the recognition of emotions by 8%.

Recently, several new computational models are reported in the literature. This includes the Neural Network for the classification of emotions/affective states from audio [Palo et al. 2015; Soltani and Ainon 2007]. Neural Network or Artificial Neural Network (ANN) consists of an input layer, one or more hidden layers and an output layer as shown in Figure 4.

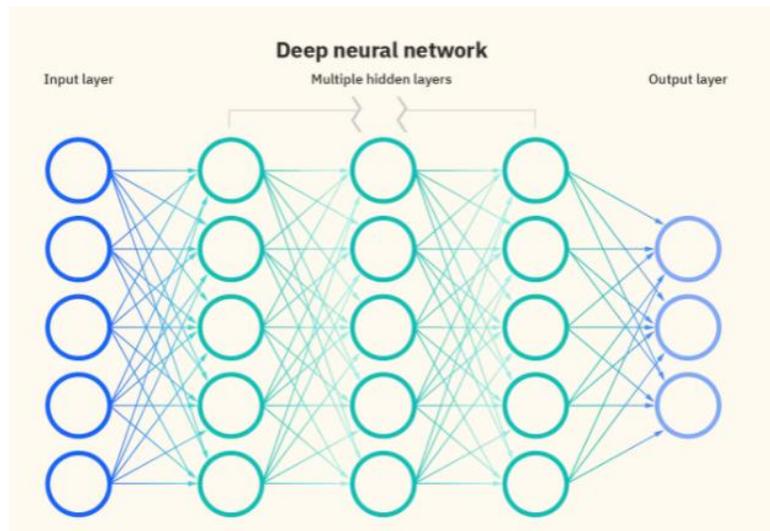

Figure 4: Neural Network or Artificial Neural Network (ANN)

The layers are made up of nodes, and each layer is connected to the next layer. Once the training of the ANN is completed with raw data, the ANN can be used to classify new data. Each node works based on its own linear regression model, which is composed of input data, weights and a threshold (bias) and an output. Machine Learning (ML) and Deep Learning (a subset of ML), which are part of Artificial Intelligence (AI), are also emerging fields for recognising affective states. Though ANN has various specific advantages over the HMM and GMM it still lacks rules for the optimal setting of ANN topology, and the learning and training makes the classification procedure difficult.

Agarwalla and Sarma used Feed-Forward (FF) network and Deep Neural Network (DNN) approaches to extract relevant data samples from big data space and used them for automatic speech recognition using soft computing techniques for Assamese speech with dialectal variations. The Feed-Forward Network is ANN where the information only moves in one direction without forming a loop between different nodes in layers. The Multi-Layer Perceptron (MLP), which is a Feed-Forward ANN, was configured with inputs to learn emotion classification information using clustering and manual labelling. The extracted features comprised spectral and prosodic features. The obtained features were applied to Recurrent Neural Network (RNN) and Fully Focused Time Delay Neural Network (FFTDNN) for the evaluation of their performance to recognise mood, dialect, speaker, and gender variations in Assamese speech. Agarwalla and Sarma tested models by considering the recognition rates which were obtained by using confusion matrices and manually calculating time. The results showed that their proposed Machine Learning (ML) based sentence extraction techniques and the composite feature set when used with RNN as a classifier outperformed all other approaches [Agarwalla and Sarma 2016]. The Recurrent Neural Network (RNN) is one of the commonly used deep learning algorithms. RNN is a family of neural networks specialised for sequential data processing and has short term internal memory. Although RNNs use training data to learn similar to other ANNs, they are distinguished as they can also learn from prior given inputs to improve current outputs. Furthermore, RNNs also share their parameters across each network layer and share the same weight within each layer of the network [IBM 2020]. On the other hand, FFTDNN is a multilayer ANN which classify pattens based on shift-invariance, without needing explicit segmentation before classification.

Various deep learning methods contribute to emotion recognition from speech or song. Deep learning involves the creation of algorithms with multiple layers, working similarly to the human brain process of sensing (message travels within many neurons before making a decision). DNN is the hidden network of algorithms as seen in the ANN's middle layers highlighted in turquoise colour in Figure 4. The input data go through different layers of DNN. DNNs then form high-level invariant appropriate features called identifiers from raw input data and then go on to classify audio and voice data for detecting emotions [Bengio et al. 2013]. Therefore, DNNs perform very well in ML tasks such as speech recognition [Dahl et al. 2011]. In recent years, the performance of the DNN has surpassed the traditional ML algorithms. Furthermore, the feature is automatically selected with the use of deep learning algorithms.

Ooi et al. [Ooi et al. 2014] used prosodic and spectral features for recognising six different emotions: anger, happiness, sadness, disgust, fear and surprise. The performance of the proposed architecture, which included two main paths, were evaluated on eNTER-FACE'05 [Martin et al. 2006] and RML database [Ryerson Multimedia Research Lab. 2017]. It revealed that the Radial Basis Functions Neural Network (RBFNN) was more accurate in recognising emotions than in comparison to RNN. RBFNN transforms the input signal to a different form, that can be fed into the single hidden layer (here called feature vector) to get linear separability.

Hao et al. [Hao, Cao, Liu, Wu and Xiao 2020] used an ensemble visual-audio emotion recognition algorithm to identify different emotions. A multi-task model (which included four sub-models) was built based on the CNN networks which is another type of DNN. The authors performed two different experiments using eNTERFACE database, showing that multi-task CNN recognised 3% over the CNN model in speaker-independent experiments and an average of 2% more over the CNN model in speaker-dependent experiments respectively. The emotion recognition accuracy rate was also reported to be increased to 81.36% in speaker-independent and 78.42% in speaker-dependent experiments.

Issa et al. [Issa et al. 2020] used one-dimensional deep CNN. The input features included MFCC, Mel-scale spectrogram, and spectral contrast features. The features were extracted using samples from the Ryerson Audio-Visual Database of Emotional Speech and Song (RAVDESS) [Livingstone and Russo 2018], Berlin (EMO-DB) [Burkhardt, Paeschke, Rolfes, Sendlmeier and Weiss 2005], and Interactive Emotional Dyadic Motion Capture (IEMOCAP) datasets [Busso et al. 2008]. They utilized an incremental process to modify initial models to improve classification accuracy. The proposed framework obtained an accuracy rate of 71.61% for RAVDESS, 86.1% for EMO-DB, 95.71% for EMO-DB and 64.3% for IE-MOCAP in speaker-independent audio classification tasks.

We summarise most common algorithms used by researchers for detecting emotions in Table 2. Furthermore, most of the research related to emotion recognition and affective state detection has emphasised detection of six universal emotions of disgust, sadness, happiness, fear, anger, and surprise. Moreover, it is quite evident from the table that a vast range of corpora such as SUSAS [Hansen et al. 1997], Emo-DB[Burkhardt, Paeschke, Rolfes, Sendlmeier and Weiss 2005] and Interactive emotional dyadic motion capture database (IEMOCAP) [Busso, Bulut, Lee, Kazemzadeh, Mower, Kim, Chang, Lee and Narayanan 2008] and Simulated Telegu Speech Corpus (IITKGP-SEHSC) [Koolagudi, Maity, Kumar, Chakrabarti and Rao 2009] are commonly utilised databases for research work.

TABLE 2

Various Models of Detection of Emotions from the Human Voice

*Anger=An, Happiness=Hp, Sadness=Sd, Boredom=Bo, Disgust=Ds, Fear=Fr, Panic=Pc, Elation=El, Shame=Sh, Pride=Pr, Despair=De, Interest=In, Contempt=Co, Neutral = Ne, Excitement = Ex, Frustration = Fu, Lombard=Lo, Loud= Lu, Surprise =Sr, Joy= Jo, Sarcastic=Sc

Here Lombard (Lo) may refer to the Lombard effect in consideration with the database used, the environment in which it was created, and how the researchers applied it in their research.

| Algorithms | Features Used | Affective States and Emotions Detected* | Performance | Reference |
|---|---|---|---|---|
| GMM | Fundamental frequency contours, pitch contours | An, Hp, Sd, Bo, Ds, Fr, Pc, El, Sh, Pr, De, In, Co | Over 77% accuracy on three different databases collected at various universities | [Busso, Lee and Narayanan 2009] |
| | Spectral features | An, Fr, Ne, Hp, Sd | 82.5% accuracy on the dataset collected by hiring participants | [Hu, Xu and Wu 2007] |
| HMM | MFCC, Pitch, TEO-CB-AUTO-ENV, 16-GA feature, 48-GA feature | Ne, An, Lm, Lo | Increased performance of emotion detection using the 16-GA features on SUSAS database | [Casale, Russo and Serrano 2007] |
| | Pitch, log energy, formant, MFCCs, | Ne, An, St, Lm, Lo | SUSAS using GSVM 90% and 92% for neutral and stress speech, 96.3% of recognition rate obtained by using HMM while 70% was obtained for 4-class style classification | [Kwon [Kwon, Chan, Hao and Lee 2003] |
| | Prosodic and spectral features | Sr, Jo, An, Fr, Ds, Sd, Ne | 70% accuracy on Spanish corpus of INTERFACE Emotional Speech Synthesis Database [Barra Chicote et al. 2008] | [Nogueiras, Moreno, Bonafonte Mariño 2001] |
| | Log frequency power coefficients (LFPC) | Lo, and Lu, St Ne | For Burmese:78.5% accuracy<br>For the Mandarin utterance: 75.7% accuracy<br>On the constructed database | [Nwe, Foo and De Silva 2003] |
| SVM | MFCC, LPCC, TEO-AutoCor | An, Ex, Fu, Hp, Ne, Sd | SVM<br>90.12% on EMO-DB, 83.2% on IEMOCAP<br><br>k-NN<br>89.3% on EMO-DB, 78% on IEMOCAP | [Bandela and Kumar 2019] |
| | Energy, pitch, MFCC coefficients, LPCC coefficients and speaker rate | An, Fr, Hp, Sd | 90.08% accuracy with Linguistic data consortium (LDC) datasets and of 65.97% with University of Georgia (UGA) datasets<br><br>MFCCs features gave an accuracy of 85.085% in comparison to LPCC features accuracy of 73.125 % | [Jain, Narayan, Balaji, Bhowmick and Muthu 2020] |
| | Fundamental frequency, formant, short-term energy, loudness and (MFCCs) Line Spectral Pairs (LPSs) | Sr, An, Fr, Ds, Hp, Sd, Ne | Accuracy reaches 81.36% and 78.42% in speaker-independent and speaker-dependent experiments conducted on eNTERFACE database [Martin, Kotsia, Macq and Pitas 2006] | [Hao, Cao, Liu, Wu and Xiao 2020] |
| | MFCCs, Spectral centroids, and MFCC derivatives | Sr, An, Fr, Ds, Hp, Sd, Ne, Sr, Bo, Ne, Ca | An accuracy of:<br>92.45% on EmoDB<br>75.69% on RAVDESS<br>and 84.11% on IITKGP-SEHSC | [Bhavan, Chauhan, Hitkul and Shah 2019] |

| | Spectral features<br>Prosodic features | An, Fr, Ds, Hp, Sd, Ne | 46.1% on LDC<br>81.3% for Berlin database of German emotional speech | [Bitouk, Verma and Nenkova 2010] |
|---|---|---|---|---|
| | Global and local Prosodic features | An, Fr, Ds, Hp, Sd, Sc, Sr | Local prosodic features performed better compared to the global prosodic features on IITKGP-SEHSC | [Rao et al. 2013] |
| | MFCC | St | 66.4% on manually collected data | [Han et al. 2018] |
| | Prosodic and spectral features | An, Hp, Sd, Ds, Fr, Su | 68.57 by using the RBF neural network | [Ooi, Seng, Ang and Chew 2014] |
| **DNN** | MFCC feature, pitch related features, | Ex, Hp, Ne, Sd | 54.3% average recognition rate | [Han et al. 2014] |
| **k-NN** | Pitch, spectral features, ZCR, Intensity, MFCCs | Hp, An, Sd, Fr and Ne | 66.24% average recognition rate on Chinese emotional data set | [Rong, Li and Chen 2009] |
| **CNN rectangular kernels** | Spectrograms | An, Hp, Sd, Bo, Ds, Fr, Ne | 80.79 % on Emo-DB and Korean Speech Database | [Badshah et al. 2019] |
| | Chroma gram, spectral contrast features, Mel-scale spectrogram, Tonnetz representation | An, Ca, Ds, Fr, Hp, Ne, Sd, Su | 71.61% for RAVDESS<br>86.1% on EMO-DB<br>95.71% for EMO-DB<br>and 64.3% for IEMOCAP with 4 classes | [Issa, Fatih Demirci and Yazici 2020] |
| **Hierarchical classifiers** | Spectral, prosodic features, mean of the log spectrum | Jo, An, Fr, Ds, Bo, Ne, Sd | An average 71.5% recognition rate on Berlin Emo-DB | [Albornoz et al. 2011] |

Few studies extended the research in the area of detecting stress during any stressful event using either non-verbal features [Han, Byun and Kang 2018; Han, Yu and Tashev 2014; Pfister and Robinson 2011] or physiological features [Barreto, Zhai and Adjouadi 2007; Gillespie et al. 2017].

Barreto et al. was one of the limited studies that detected stress by monitoring physiological signals, such as galvanic skin response, temperature of skin, blood pulse volume, and left eye pupil diameter of the subjects, during a determined computer-based, paced Stroop test. The Stroop incongruent segment was considered to be related to the stressed state in participants, while congruent Stroop was linked to a non-stressed state in participants. Twelve out of 32 samples were used to train the classifiers while the remaining 20 samples were used to test the classifiers. All the extracted features of the physiological signals, such as the mean amplitude of the individual blood pulse volume beat and the mean value of pupil diameter, which were expected to increase under stress, were given as the input to various machine learning algorithms, such as Naïve Bayes, Decision Tree, and SVM classifiers, to classify the relaxed versus stressed state in participants [Barreto, Zhai and Adjouadi 2007].

Another study conducted by Han et al. proposed an algorithm to determine the state of stress (stressed or unstressed) via voice features using a binary decision criterion involving two layers of Long Short-Term Memory Recurrent Neural Networks (LSTM-RNN) and a classifier. The authors evaluated the proposed algorithm by obtaining the speech, video, and bio-signal data under stressful (interview of the Korean speaking subjects in English by a foreigner) and non-stressful (comfortable conditions such as watching videos) conditions and surveyed participants about the stress they felt. Based on the survey results and cortisol levels, the speech signals from 25 subjects whose salivary cortisol level changed more than 10% were labelled as stress or non-stress in the database. Additionally, 15 subjects were used to train the model, while five stressed and five un-stressed speech samples from the database were used to test the data. The authors pre-processed the data before extracting the MFCC, which were used to determine stressed and unstressed state from the testing data. The proposed LSTM-RNN with SVM classifier achieved 66.4% accuracy in stress detection [Han, Byun and Kang 2018].

Furthermore, Pfister and Robinson presented a new classification algorithm to assess public speaking skills. They trained the classifier using the Mind Reading corpus [Junek 2007] for the detection of both simple and complex emotions. For the classifier to detect the emotions, the authors chose a wide variety of emotions such as joy, interestedness, unfriendliness, excitement, unsureness, and sureness from the Mind Reading corpus. They pre-processed the corpora, extracted the non-verbal features of speech and computed the SVM model. Then, to assess public speaking skills, Pfister and Robinson first retrained the classifier using six labels which are commonly used by professional experts. For this purpose, they asked an experienced speech coach to label 124 one-minute-long speech samples obtained from 31 speakers attending speech coaching. Finally, Pfister and Robinson segmented the live audio, extracted the audio features, and ran the classifiers. For pairwise machine, the results yielded an average cross-validation accuracy of 89% while for the fused machine, 86% of accuracy was obtained. For assessing public speaking skills, the novel application of the classifier achieved 81% cross-validation accuracy and 61% accuracy when performed using a leave-one-speaker-out method [Pfister and Robinson 2011].

Specifically, stress is common during public speaking. Although public speaking is recognised as an important skill, it has received little attention for real-time voice analysis. This is most possibly due to limitations and complexities involved in creating a suitable testing environment. To overcome this, technologies like VR will be more helpful in creating a public speaking simulation.

## 4 VIRTUAL REALITY FOR PUBLIC SPEAKING

VR has also been utilised for improving public speaking skills [Kimani and Bickmore 2019; Lister et al. 2010; North et al. 1998; Takac et al. 2019], and other interpersonal skills. The benefit of the VR environment is the immersive experience it brings to users, which provides the experience of learning in a way that is like real-life learning [Azwar, Alam, Kazmi, Zain-ul-abidin and khan 2016; Gavish et al. 2015; Nijholt 2014].Therefore, learning skills using VR is becoming increasingly popular as well [Dascalu et al. 2017]. It is therefore useful in prototyping, scientific visualization, training, engineering, manufacturing, and learning [Dillon et al. 2006; Kassem et al. 2017; Le et al. 2015; Pedro et al. 2016; Schmid Mast et al. 2018].

Researchers have applied VR in various settings, and some of them are summarized in Table 3 [Bhagat et al. 2016; Bouchlaghem et al. 2005; Carl et al. 2019; Chittaro and Sioni 2015; Crocetta et al. 2018; Dechant et al. 2017; Garcia-Palacios, Hoffman, Carlin, Furness and Botella 2002; Gebara et al. 2016; Gerardi et al. 2010; Hilfert and König 2016; Kim et al. 2009; Laver et al. 2015; Levy et al. 2016; Lindner et al. 2019; Manju et al. 2017; Maskey et al. 2014; Miloff et al. 2016; Molina et al. 2014; Motraghi et al. 2014; Mujber et al. 2004; Opriş et al. 2012; Ordaz et al. 2015; Parsons and Rizzo 2008; Ticknor 2018; Ticknor 2019; Valmaggia et al. 2016].

TABLE 3
Various Applications of Virtual Reality (VR)

| Category | Broad Applications | Specific applications |
|---|---|---|
| **Industrial** | Gaming | Stroke Rehabilitation |
| | Architectural | |
| | Design Construction | |
| | Manufacturing | |
| | Healthcare | Diagnostic tool |
| **Educational** | Learning | Social Skills Training |
| **General** | Overcoming fears | Fear of heights |
| | Relaxation therapies | Fear of spiders |
| | | Fear of public speaking |
| **Clinical Psychology** | Assessing & treating disorders | Anxiety, Posttraumatic Stress Disorder |
| | Complement to standard CBT treatments | Schizophrenia |
| | | Depression |
| | | Eating disorders |

Furthermore, there has been much research interest in using VR to explore feedback strategies to provide social support in a VR environment for public speaking [Brundage and Hancock 2015; Chollet et al. 2015]. It is reported that psychosocial stress due to public speaking is a common occurrence across the world, which may hinder the quality of the speech delivered [Koroleva, Bakhchina, Shyshalov, Parin and Polevaia 2014; Ruscio et al. 2008]. Studies suggest that interactions with virtual entities and their behaviours may be able to elicit both social stress and social support [Kothgassner et al. 2016; Kothgassner et al. 2019; Kotlyar, Donahue, Thuras, Kushner, O'Gorman, Smith and Adson 2008; Pan et al. 2012]. Conducting public speaking tasks in a laboratory setting can help in the evaluation of real-life stress and finding ways to overcome it as well [Feldman et al. 2004; Jezova et al. 2016; Owens et al. 2015].

Brundage & Hancock in 2015 investigated VR exposure therapy for treating people who stutter. In their study, ten people who stutter first delivered a speech to a real audience and then, on another day, performed in front of two virtual audience members. The study results indicated that participants had an experience in the virtual setting similar to their experience with the real audience [Brundage and Hancock 2015]. Therefore, VR can be used as a standardised tool for stress [Jönsson et al. 2010] and fear [Powers et al. 2013] induction in real life. Notably, these studies show how people around us, even in a virtual setting, can influence our feelings, behaviour, and performance. In another research study, the authors explored feedback strategies for public speaking by training a virtual audience to be non-interactive (control condition), or to provide direct visual feedback, or non-verbal feedback from non-verbal behaviour such as eye contact and pause fillers [Chollet et al. 2015; Wörtwein et al. 2015].

Furthermore, research on VR exposure therapy was investigated by Hartanto et al. [Hartanto et al. 2014] to be used for the treatment of anxiety disorders by controlling social stress. The intervention included a one-to-one dialogue session between the speaker and the virtual audience. The participants were exposed to three virtual scenarios of a neutral virtual scene, a blind date condition and a job interview session. The participants' anxiety levels were significantly increased when the social situation changed.

In one of the earliest studies conducted by Slater et al.1999, studied the psychotherapeutic effects of a virtual environment on social phobia for public speaking. The study had an experimental design of a virtual public scenario with avatars. The avatars displayed random behaviours such as nodding, twitching, and blinking. Avatars could also yawn, clap, and show hostile reactions. Ten graduate students were recruited from University College London (UCL). The participants either gave their speech to the audience on the monitor or were entirely 'immersed' in a VR setting with a headset. Each participant repeated their speech three times. At the first time, the participants faced either a hostile or a friendly audience. At the second time, participants faced whichever audience they did not face during the first speech. At the third time, participants faced an audience with hostile reactions, which eventually become friendly. A participant immersed using the VR headset reported a high co-presence with a hostile audience (felt as though the participant was together with the listeners in real-life) but did not feel as though the audience was interested. This shows that co-presence amplified the situation and a low-interest audience reduced self-rating and increased anxiety in the speaker [Slater, Pertaub and Steed 1999]

In a separate study by Slater et al. 2006, therapeutic intervention in VR was used to reduce the fear of public speaking. The experiment included 16 participants with a fear of public speaking and 20 participants with a lower amount of fear of public speaking to give a speech in an empty seminar room in a VR setting. Later, the same room was populated by a neutrally behaving virtual audience of five people. The responses from participants with a fear of public speaking showed a significant increase in signs of anxiety when speaking to the virtual audience in comparison to when speaking to the empty room. In contrast, the participants with a lower amount of fear of public speaking reported no anxiety at all [Slater, Pertaub, Barker and Clark 2006].

Significant differences were detected between groups undergoing virtual reality therapy (VRT) sessions and a control group, with VRT found to be effective in reducing the fear of public speaking [North et al. 1998]. Furthermore, another study showed that a hostile, negative audience scenario generated a strong effect in speakers, regardless of whether the participants feared public speaking. The results suggested that induced anxiety is directly related to the virtual audience's feedback [Pertaub et al. 2001].

A group of researchers developed a virtual character system, which accepts an audio signal's transcription as the input [Pellett and Zaidi 2019]. Based on the analysis of the audio signal received, the system generates an animated performance as output. A framework for virtual reality training to improve public speaking was proposed, which involved assessing the speech using a speech-to-text approach [Marsella et al. 2013], using the Windows dictation recogniser. The speech performance results were shown to speakers at the end of their speech to grade their overall performance. Another more inventive presentation trainer was developed to assess the public speaking performance by giving the user real-time feedback about different aspects of the participant's non-verbal communication. It tracks the participant's voice and body to interpret performance and present an intervention to give feedback to the user [Schneider et al. 2015]. In another research study, a group of researchers developed an interface using Google Glass to help people to improve public speaking by automatically detecting the speaker's volume and speaking rate in real time. Based on the speaker's volume and speaking rate, the system provides feedback during the speech [Tanveer et al. 2015].

Similarly, El-Yamri et al. 2019, created a VR system, which included a virtual audience character model. Emotions such as stress elicited in the speaker by the virtual audience character model were analysed using a third-party emotion detection application. The detected emotional states, such as stress and happiness, were weighted using the authors' common sense only. This may not be accurate due to insufficient and unclear information for the basis of common sense used. Moreover, this model lacks a learning process for the virtual agent based on the analysis of voice features and other factors such as physiological parameters. For example, the authors could have detected emotions using VR in real time and/or integrating the developed VR application with another built emotion detection model to analyse voice features and physiological parameters. Based on the detected emotions, the authors could have then given the score to the speaker. This score could have been used to train the virtual audience to provide real-time feedback to the speaker based on a certain score of speakers [El-Yamri et al. 2019].

Table 4 shows some of the relevant studies on public speaking in VR.

TABLE 4
Various Relevant Studies of Public Speaking in Virtual Reality (VR)

| Article | Main issue focused on | Task | Responses of virtual audience |
|---|---|---|---|
| [Brundage and Hancock 2015] | Affect generation | Speech | Neutral and challenging virtual audience |
| [Chollet, Wörtwein, Morency, Shapiro and Scherer 2015] | Anxiety | Speech | Non-verbal indirect feedback |
| [El-Yamri, Romero-Hernandez, Gonzalez-Riojo and Manero 2019] | Fear/stress | Speech | Real-time feedback voice tone, gaze, and speech content |
| [Kothgassner, Felnhofer, Hlavacs, Beutl, Palme, Kryspin-Exner and Glenk 2016; Kothgassner, Goreis, Kafka, Kaufmann, Atteneder, Beutl, Hennig-Fast, Hlavacs and Felnhofer 2019] | Stress | Speech | Real audience, Virtual audience, Empty virtual lecture hall |
| [Pertaub, Slater and Barker 2001] | Fear | Speech | Positive, negative comments and behavior |
| [Slater, Pertaub, Barker and Clark 2006; Slater, Pertaub and Steed 1999] | Anxiety | Speech | Positive and negative evaluation feedback |
| [Hartanto, Kampmann, Morina, Emmelkamp, Neerincx and Brinkman 2014] | Stress | One-to-one conversation session | Neutral and conversing audience |

Overall, most studies related to public speaking in VR show a lack of clarity in terms of how the affective state of the speaker was determined in real time during the tasks. Furthermore, changes in the speaker's affective state due to the virtual audience's feedback is or was solely understood based either only on self-reported questionnaires or physiological change. An additional way to determine speaker's affective state is to assess emotions in real time by analysing the voice. Interestingly, emotion/affective state detection from voice has unique traits and also common elements shared with the analysis of emotional contents in music, as discussed in one of earlier works by Scherer [Scherer 1995]. Particularly for detection of stress from voice, a approach similar to discussed by Dillon in research conducted can also be investigated [Dillon 2001; Dillon 2003].

In next Section 5 we will discuss the gaps, opportunities, potential barriers, and suggestions based on the literature survey in more details.

## 5 DISCUSSION

It is well-known and established that communication by speech is a powerful means of human expression and connection. The human voice has proven to be a good indicator of emotional/affective state in several studies [Cowen, Elfenbein, Laukka and Keltner 2019; Juslin and Laukka 2003; Mitchell and Ross 2013]. In the areas of HCI and affective computing, recognition of emotions and human affective states is one of the most challenging tasks. Emotions such as anxiety and affective states such as stress are subjective to individuals. Therefore, the development of a detection model depends on the affective state causing event and its application area.

Various models exist and are utilised for analysing the human voice to detect emotions/affective states. Some studies suggested that prosodic features work more appropriately in the identification of emotion with the GMM [Li, Tao, Johnson, Soltis, Savage, Leong and Newman 2007]. Previous research also suggests that the HMM approach achieves a higher recognition rate when detecting happiness, anger, joy, fear, disgust, and sadness using spectral features such as MFCC and LPCC [Nogueiras, Moreno, Bonafonte and Mariño 2001; Nwe, Foo and De Silva 2003]. Furthermore, in a comparison of HMM and SVM to classify five emotions (anger, happiness, sadness, surprise, and neutral emotion), the results showed a 99.5% accuracy rate for HMM, whereas SVM obtained an accuracy rate of only 88.9% in recognizing the emotions [Lin and Wei 2005]. However, other research studies indicate SVM as promising for the classification of stress or not stressed or other states [Barreto, Zhai and Adjouadi 2007; Truong et al. 2012]. Other computational models such as Artificial Neural Network (ANN), Recurrent Neural Network (RNN), and Convolutional Neural Network (CNN) are also reported to perform well. However, in recent years, the performance of the Deep Neural Network (DNN) has surpassed the traditional Machine Learning (ML) algorithm [Dahl, Yu, Deng and Acero 2011].

Most of the research related to emotion recognition and affective state detection has emphasized the detection of six universal emotions (i.e., disgust, sadness, happiness, fear, anger, and surprise) using spectral, temporal, and prosodic features [Bhavan, Chauhan, Hitkul and Shah 2019; Busso, Lee and Narayanan 2009; Hao, Cao, Liu, Wu and Xiao 2020; Hu, Xu and Wu 2007; Jain, Narayan, Balaji, Bhowmick and Muthu 2020; Nogueiras, Moreno, Bonafonte and Mariño 2001; Nwe, Foo and De Silva 2003; Ooi, Seng, Ang and Chew 2014; Schuller et al. 2003; Semwal, Kumar and Narayanan 2017]. Furthermore, the extraction of voice features depends upon emotion or affective state that needs identification. For example, Low et al. [Low, Maddage, Lech, Sheeber and Allen 2010] concluded that specifically for the identification of stress, Teager Energy Operators (TEO) based features, and especially the TEO CB AutoEnv, are more effective in comparison to all other types of features. Furthermore, sometimes the voice features may overlap between different emotions, and therefore, it is crucial to extract correct features carefully.

From these studies, it can be inferred that various voice features, and various datasets and models play a different role in emotion recognition and affective state identification. It depends on the dataset we choose the emotion we want to detect, and the preference of algorithm. A combination of carefully extracted features, with a proper noise-free training data set and with the appropriate classification algorithm, is needed to develop an accurate detection model. Once the application of emotion detection is validated within a specific context, it is possible to port it into other settings to detect stressful events of different nature with an improved recognition rate.

**5.1 Gaps and Opportunities**

Public speaking is recognised as an essential skill as well as a cause of stress. Acquiring public speaking skills for the presentation of ideas is important in various professional contexts [Docan-Morgan and Nelson 2015]. It is reported that psychosocial stress due to public speaking is a common occurrence across the world, which hinders the quality of the speech delivered [Koroleva, Bakhchina, Shyshalov, Parin and Polevaia 2014; Ruscio, Brown, Chiu, Sareen, Stein and Kessler 2008]. Although much recent research has been done in the field of emotion detection, not enough research has been carried out to specifically detect stress in the context of a potentially stressful event such as public speaking in real time.

More specifically, public speaking has received little attention for real-time stress detection by analysing voice features and physiological parameters. Very few studies have extended research to the matter of detecting stress during any potentially stressful event using either non-verbal features [Han, Byun and Kang 2018; Pfister and Robinson 2011] or physiological features [Barreto, Zhai and Adjouadi 2007; Gillespie, Moore, Laures-Gore, Farina, Russell, Logan and Ieee 2017] or self-reported indices [Bandela and Kumar 2019].

For example, Pfister and Robinson [2011] utilised non-verbal features to detect the affective state and apply it to assess public speaking only, leaving out the physiological parameters. The accuracy of this method of detecting stress may not be high because of the restriction to just non-verbal features, leaving out the physiological parameters in the context of public speaking.

In another example, in the study conducted by El-Yamri et al. [El-Yamri, Romero-Hernandez, Gonzalez-Riojo and Manero 2019] where the authors created a VR system, which included a virtual audience character model. Stress elicited in a speaker was identified and analysed using a third-party affective state detection application. Furthermore, the emotional states of the speaker were weighted using the authors' common sense only. This is not an appropriate and accurate option due to insufficient and unclear information for the basis of 'common sense' used, which can be culturally dependent and hence could potentially bias the recognition of certain emotions over others. Moreover, this model lacks a learning process for the virtual agent based on the analysis of voice features and other factors such as physiological features. Also, knowing how the virtual audience were made to work can help in improving the experiment and real-time feedback to the speaker during public speaking.

The literature review points out that real-time stress detection in the areas of affective computing and HCI has not received much attention, and this is possibly due to the limitations and complexities in creating a suitable testing environment for stress during public speaking. While conducting public speaking tasks in a laboratory setting can help in the evaluation of real-life stress [Jezova, Hlavacova, Dicko, Solarikova and Brezina 2016], technologies like VR can be employed to create a simulation for public speaking. Previous studies have also indicated that interactions with virtual entities are able to elicit social stress. It can also be inferred from research that VR can be used to provide Virtual Social Support (VSS) during a stressful activity [Kothgassner, Goreis, Kafka, Kaufmann, Atteneder, Beutl, Hennig-Fast, Hlavacs and Felnhofer 2019]. In this regard, research has been carried out using VR to examine the effects of VSS for public speaking [Chollet, Stefanov, Prendinger and Scherer 2015; Chollet, Wörtwein, Morency, Shapiro and Scherer 2015; El-Yamri, Romero-Hernandez, Gonzalez-Riojo and Manero 2019; Hartanto, Kampmann, Morina, Emmelkamp, Neerincx and Brinkman 2014; Pertaub, Slater and Barker 2001; Slater, Pertaub, Barker and Clark 2006; Slater, Pertaub and Steed 1999]. However, most of the research (e.g., [Barreto, Zhai and Adjouadi 2007; Hartanto, Kampmann, Morina, Emmelkamp, Neerincx and Brinkman 2014; Slater, Pertaub, Barker and Clark 2006]) related to VR for public speaking focused on testing the effects of virtual audience behaviour on a speaker's anxiety and fear of public speaking by only measuring the speaker's physiological parameters such as heart rate or blood pressure. Despite being important indices to emotional experience, changes in heart rate and blood pressure are not exclusively indicative of specific emotional experience. The research could have been improved by detecting emotions from speakers' voices, which could have made the detection of stress more accurate since the human voice is a powerful channel to convey emotions and this has not been done using VR before as accurately as it could be done.

Based on the gaps identified, we propose the development of a stress detection model. This model would be able to analyse the voice of users in real time. The selected voice features to develop the stress detection model will be correlated to the physiological parameters of stress such as heart rate and blood pressure as well as psychological results obtained from self-reported questionnaires such as stress arousal checklist [Cox and Mackay 1985] and the task specific anxiety checklist.

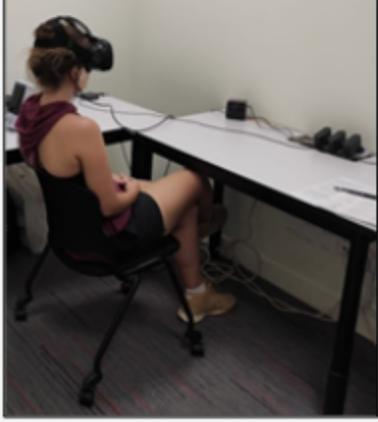

Figure 5: Data collection in process for developing stress detection model

After data collection (Figure 6), the first step would be to analyse the self-reported responses in pre- and post-test questionnaires for each condition for all speakers. We will then check the corresponding heart-rate readings and blood pressure readings of the participants for both VR scenarios. If both the psychological responses and physiological responses indicate that the person was stressed during public speaking in VR, we will proceed to extract the voice features, which will be counted as our stressed voice data. On the other hand, the speeches provided by participants who did not show signs of stress during their speech will be kept as our confident voice data set. Once the features of voice are extracted, we will evaluate various algorithms and choose the one that can provide us with the best accuracy for developing the stress detection model.

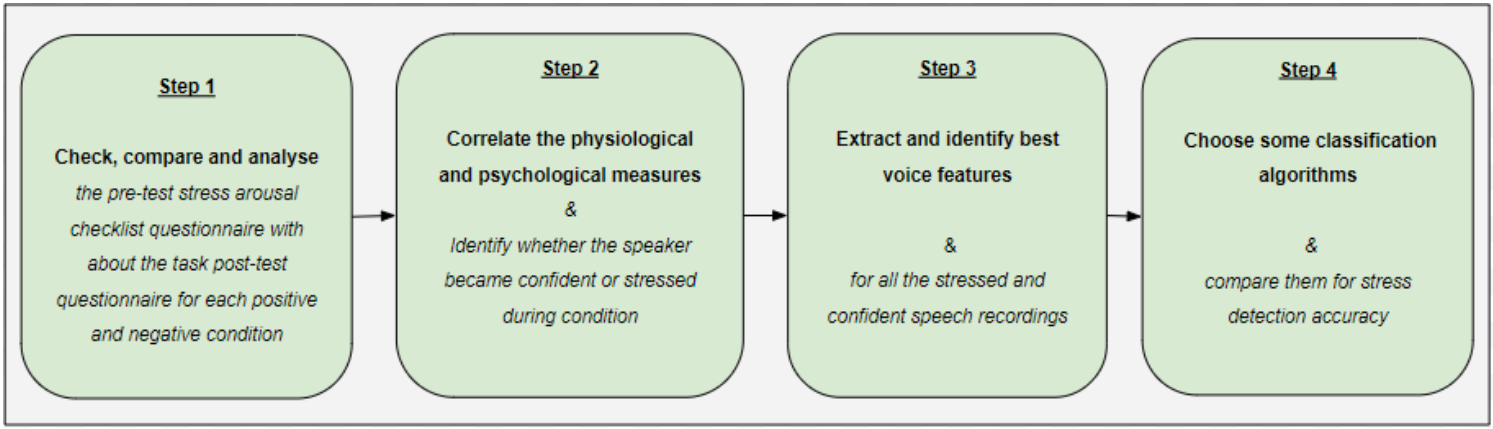

Figure 6: Steps in Development of Stress Detection Model to be Integrated in VR application

The developed stress detection model will then be integrated or connected to a virtual audience in the VR application to provide real-time feedback in two modes (social supportive and realistic feedback mode) (Figure 7).

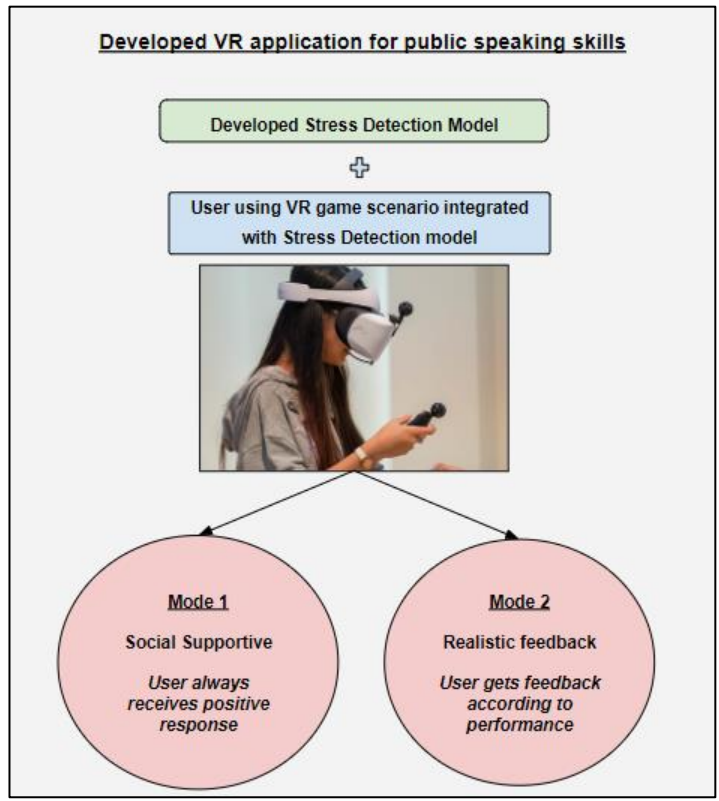

Figure 7: VR game application for improving public speaking skills

This would allow users to practice their public speaking in social supportive mode (when less certain about their speaking abilities) or they could choose to practice in realistic feedback mode (when sure about their public speaking skills and wanting to check their performance). This will allow users to gradually learn to overcome stress caused by public speaking and therefore improve their public speaking skills.

### 5.2 Potential Barriers and Solutions

Several issues may arise in the implementation procedure of the suggestion presented in 5.1. For example, there may be barriers in the development software for extraction and analysis of all necessary voice features. Issues may also arise if we want to analyse the physiological parameters of stress in real time. We can start by pre-processing of recorded voice data, followed by extracting the spectral features of voice such as frequency contours, and prosodic features using Python libraries like My Voice Analysis [Shabahi 2020], LibROSA [McFee et al. 2015] or pyAudioAnalysis [Giannakopoulos 2015]. Furthermore, if needed, similar to approach labelled by Batrinca, we can also seek expert advice in deciding various prosodic features [Batrinca et al. 2013]

In case issues arise with the analysis of the physiological parameters of stress in real time, we can also develop another application to record the physiological parameters separately for analysis of stress levels. We may then try to combine the results within the emotion detection model by training it with voice features combined with physiological parameters specifically for detecting stress. This would allow the integration of the whole developed model into the VR application using TensorFlow. The built emotion detection model would be based on analysis of voice features such as prosodic features, spectral features, and physiological parameters of stress.

This model could be used to train the virtual audience present in the VR application using the Machine Learning Agents SDK (ML-Agents) using various machine learning methods such as reinforcement learning. This would allow the virtual agents to provide real-time feedback to the speaker with respect to voice and physiological features.

Therefore, analysis of stress elicited in a public speaker by a virtual audience should be accomplished by using stress detection models in VR. Furthermore, the developed stress detection model combined with voice analysis and physiological symptoms of stress could be used to train the virtual audience in real time to provide feedback. From the research surveyed here, it is evident that stress detection models in VR have not yet been utilised for providing real-time feedback based on both voice analysis and physiological symptoms.

## 6 CONCLUSION

Results and findings from the present review are encouraging. There are opportunities to advance the research in voice analysis for specifically detection of stress in a VR setup during public speaking. Limited research has been carried out to detect the real-time presence of stress during a stressful event like public speaking. This is likely due to the complexities and limitations involved in creating a suitable environment. Exponential technologies such as VR can help provide an immersive environment for this very purpose. Therefore, the detection of stress from the voice should be explored by developing a stress-detection model and implementing it in VR. However, the potential barriers include analysis of prosodic features of voice and especially real-time analysis of physiological parameters of stress during public speaking in VR. The issues can be overcome by either analysing physiological signals within the VR application or by using another application and then importing the physiological data into the VR application. This review has its limitations such as in terms of limited number of articles surveyed. However, it shows that there are opportunities for integrating real-time audio and voice analysis within VR, which can help detect stress during public speaking and, ultimately, provide a valid training tool to improve public speaking skills by the suggested idea.

Elfering, A. and Grebner, S. 2011. Ambulatory Assessment of Skin Conductivity During First Thesis Presentation: Lower Self-Confidence Predicts Prolonged Stress Response. *Applied Psychophysiology and Biofeedback 36*, 93-99.

Elfering, A. and Grebner, S. 2012. Getting Used to Academic Public Speaking: Global Self-Esteem Predicts Habituation in Blood Pressure Response to Repeated Thesis Presentations. *Applied Psychophysiology and Biofeedback 37*, 109-120.

Feldman, P.J., Cohen, S., Hamrick, N. and Lepore, S.J. 2004. Psychological Stress, Appraisal, Emotion And Cardiovascular Response In A Public Speaking Task. *Psychology & Health 19*, 353-368.

Fredrick, S.S., Demaray, M.K., Malecki, C.K. and Dorio, N.B. 2018. Can social support buffer the association between depression and suicidal ideation in adolescent boys and girls? *Psychology in the Schools 55*, 490-505.

Fruchter, R., Reidsma, D., Op Den Akker, H.J.A., Nishida, T., Rienks, R.J., Rosenberg, D., Poppe, R.W., Nijholt, A., Heylen, D.K.J. and Zwiers, J. 2007. Virtual Meeting Rooms: From Observation to Simulation. *AI & society 22*, 133-144.

Gandhi, R. 2018. Support Vector Machine — Introduction to Machine Learning Algorithms.

Garcia-Palacios, A., Hoffman, H., Carlin, A., Furness, T.A. and Botella, C. 2002. Virtual Reality In The Treatment Of Spider Phobia: A Controlled Study. *Behaviour research and therapy 40*, 983-993.

Gavish, N., Gutiérrez, T., Webel, S., Rodríguez, J., Peveri, M., Bockholt, U. and Tecchia, F. 2015. Evaluating Virtual Reality And Augmented Reality Training For Industrial Maintenance And Assembly Tasks. *Interactive Learning Environments 23*, 778-798.

Gebara, C.M., Barros-Neto, T.P.D., Gertsenchtein, L. and Lotufo-Neto, F. 2016. Virtual Reality Exposure Using Three-Dimensional Images For The Treatment Of Social Phobia. *Revista brasileira de psiquiatria (Sao Paulo, Brazil : 1999) 38*, 24-29.

Georgiades, A., Sherwood, A., Gullette, E.C.D., Babyak, M.A., Hinderliter, A., Waugh, R., Tweedy, D., Craighead, L., Bloomer, R. and Blumenthal, J.A. 2000. Effects of Exercise and Weight Loss on Mental Stress–Induced Cardiovascular Responses in Individuals With High Blood Pressure. *Hypertension: Journal of the American Heart Association 36*, 171-176.

Gerardi, M., Cukor, J., Difede, J., Rizzo, A. and Rothbaum, B.O. 2010. Virtual Reality Exposure Therapy for Post-Traumatic Stress Disorder and Other Anxiety Disorders. *Current Psychiatry Reports 12*, 298-305.

Giannakopoulos, T. 2015. pyAudioAnalysis: An Open-Source Python Library for Audio Signal Analysis. *PloS one 10*, e0144610.
31

Martin., D.J.J.H. 2021. Hidden Markov Models. In *Speech and Language Processing*.

Maskey, M., Lowry, J., Rodgers, J., Mcconachie, H. and Parr, J.R. 2014. Reducing specific phobia/fear in young people with autism spectrum disorders (ASDs) through a virtual reality environment intervention. *PloS one 9*, e100374.

Mcfee, B., Raffel, C., Liang, D., Ellis, D.P., Mcvicar, M., Battenberg, E. and Nieto, O. 2015. librosa: Audio and music signal analysis in python. In *Proceedings of the 14th python in science conference*, 18-25.

Miloff, A., Lindner, P., Hamilton, W., Reuterskiöld, L., Andersson, G., Carlbring, P., Stockholms, U., Samhällsvetenskapliga, F., Psykologiska, I. and Klinisk, P. 2016. Single-session gamified virtual reality exposure therapy for spider phobia vs. traditional exposure therapy: study protocol for a randomized controlled non-inferiority trial. *Trials 17*, 60.

Mitchell, R.L. and Ross, E.D. 2013. Attitudinal Prosody: What We Know And Directions For Future Study. *Neuroscience & Biobehavioral Reviews 37*, 471-479.

Molina, K.I., Ricci, N.A., De Moraes, S.A. and Perracini, M.R. 2014. Virtual reality using games for improving physical functioning in older adults: a systematic review. *Journal of NeuroEngineering and Rehabilitation 11*, 156.

Motraghi, T.E., Seim, R.W., Meyer, E.C. and Morissette, S.B. 2014. Virtual Reality Exposure Therapy for the Treatment of Posttraumatic Stress Disorder: A Methodological Review Using CONSORT Guidelines. *Journal of Clinical Psychology 70*, 197-208.

Mujber, T.S., Szecsi, T. and Hashmi, M.S. 2004. Virtual Reality Applications In Manufacturing Process Simulation. *Journal of materials processing technology 155*, 1834-1838.

Neiberg, D., Elenius, K., Laskowski, K. and Isca 2006. *Emotion Recognition in Spontaneous Speech Using GMMs*.

Nijholt, A. 2014. Breaking Fresh Ground in Human-Media Interaction Research. *Frontiers in ICT 2014*.

Nogueiras, A., Moreno, A., Bonafonte, A. and Mariño, J.B. 2001. Speech emotion recognition using hidden Markov models. In *Seventh European Conference on Speech Communication and Technology*.

North, M.M., North, S.M. and Coble, J.R. 1998. Virtual Reality Therapy: an Effective Treatment for the Fear of Public Speaking. *International Journal of Virtual Reality 3*, 1-6.
36

41